\documentclass[a4paper,english,11pt,bibliography=totoc]{article}

\usepackage[T1]{fontenc}
\usepackage[utf8]{inputenc}	
\sffamily		
\usepackage[top=2cm, bottom=2cm, left=2cm, right=2cm]{geometry}				
\usepackage{amsmath}			
\usepackage{siunitx}			
\usepackage{amsmath}
\usepackage{amsfonts}
\usepackage{amssymb}
\usepackage{graphicx}			
\usepackage[round]{natbib}		
\usepackage{hyperref}			
\usepackage{float}
\usepackage{color}
\usepackage{paralist}
\usepackage{url}
\usepackage{hyperref}

\makeatletter 
\newcommand*\bigcdot{\mathpalette\bigcdot@{.7}}
\newcommand*\bigcdot@[2]{\mathbin{\vcenter{\hbox{\scalebox{#2}{$\m@th#1\bullet$}}}}}
\makeatother

\usepackage{helvet}

\usepackage{titlesec}
\titleformat{\section}
{\normalfont\sffamily\Large\bfseries\color{black}}
{\thesection}{1em}{}
\titleformat{\subsection}
{\normalfont\sffamily\large\bfseries\color{black}}
{\thesubsection}{0.6em}{}

\usepackage[font={footnotesize,it}]{caption}

\begin{document}
\fontfamily{phv}\selectfont

\vspace{-1.8cm}
\begin{center}
	\LARGE{\textbf{Reddy: An open-source toolbox for analyzing eddy-covariance measurements in heterogeneous environments}} \\[0.6cm]
	
	\normalsize
	Laura Mack$^{1,*}$, Norbert Pirk$^1$ \\[0.2cm]
	
	\footnotesize
	1: Department of Geosciences, University of Oslo, Oslo, Norway \\
	*Correspondence: laura.mack@geo.uio.no
\end{center}

\normalsize

\section*{Abstract}	
	Land-atmosphere exchange is mediated by turbulent fluxes that can be quantified using eddy-co\-variance (EC) measurements. EC has been widely used to measure ecosystem-scale vertical exchange between atmosphere and vegetation, and to test and refine atmospheric turbulence theories with the aim to improve the representation of turbulent fluxes in numerical models. Traditionally, research has focused on idealized, homogeneous and flat surfaces, but recent work increasingly targets turbulent exchange in complex, heterogeneous environments under non-ideal conditions, where challenges include advective fluxes, mesoscale circulations between contrasting surface types, and non-stationary nighttime turbulence.		
	Here, we introduce the open-source R package Reddy, which combines multiple EC analysis methods into a single modular tool. Reddy enables users to tailor post-processing choices to site-specific conditions, supports station management and facilitates detailed scientific analyses. The package is accompanied by extensive documentation and a suite of Jupyter notebooks that provide hands-on introductions to EC data processing.		
	We demonstrate Reddy using measurements from three Norwegian sites: (1) a morning transition following a strongly stably stratified night at an alpine tundra valley, (2) spectral and ogive analysis before and after an ice-cover transition at a boreal lake, and (3) fitting flux-variance relations at a permafrost-affected palsa peatland. Reddy extends existing EC software and helps moving towards a more holistic turbulence data analysis framework for heterogeneous, real-world environments.\\
	

	\textbf{Keywords:} Boreal Lake, Heterogeneous Terrain, Land-Atmosphere Exchange, Open-Source Software, Turbulent Fluxes

\section{Introduction}
Turbulent fluxes are important drivers of land-atmosphere and vegetation-atmosphere interactions \citep[e.g.][]{Baldocchi2003,Baldocchi2020}, and the standard technique for measuring them is eddy-covariance (EC) \citep[e.g.][]{Aubinet2012}. Strictly speaking, however, EC is only the method used to calculate turbulent fluxes from high-frequency ultra-sonic anemometer and gas analyzer measurements of three-dimensional wind speed, temperature and trace gases, which also enable in-depth analysis of turbulence characteristics. 
To derive turbulent fluxes of sensible heat, latent heat, carbon dioxide or other trace gases from the high-frequency measurements, several software packages have been developed that offer a range of (post-)processing options for deriving vertical exchange between vegetation and atmosphere, most notably EddyPro \citep{EddyPro2017}.
These software packages are well-tested and robust, leading to a high agreement of the derived turbulent fluxes \citep{Mauder2008,Fratini2014,Mammarella2016}, and have served as tools to generate regional or global flux products, e.g. within FLUXNET \citep{Pastorello2020} or the Integrated Carbon Observation System (ICOS). 
However, recent scientific developments are moving towards studying turbulent fluxes in more complex heterogeneous environments \citep[e.g.][]{Rotach2025,Stiperski2025}, which requires adapting common post-processing routines that were originally developed for homogeneous flat terrain under idealized well-mixed turbulence conditions -- as briefly summarized below.

\subsection{Limitations of the EC method in heterogeneous environments}
The EC method involves several sources of uncertainties in general, which are often amplified at heterogeneous sites \citep[e.g.][]{Loescher2006,Mauder2006,Mauder2008,Mauder2013}. These uncertainties can be roughly grouped into (1) conceptual/theoretical, (2) instrumental, (3) data-processing, and (4) site- or surface-dependent limitations: 

\begin{itemize}
	\item \textbf{Conceptual limitations:} The theoretical foundation of the EC method assumes horizontal homogeneity and stationarity (i.e. that the statistical turbulence properties are constant over the chosen averaging time), and typically neglects advection and storage terms. In heterogeneous environments, surface properties can change abruptly (e.g. snow patches, lake--forest interfaces), so that the EC flux represents a mixture of flux contributions from different surface types. 
	Horizontally varying surface properties can induce horizontal advection and mesoscale circulations, so that a one-dimensional EC setup does not represent the full three-dimensional budget. The EC method requires well-developed turbulence, which is not necessarily fulfilled at night when stable stratification suppresses turbulence.
	
	\item \textbf{Instrumental limitations:} The EC setup is subject to technical limitations, including limited sensor response times leading to high- and low-frequency losses, sensor separation and imperfect alignment causing phase shifts and flux attenuation, and calibration uncertainties, in particular for gas analyzers. While these are intrinsic to EC in general, measurements in heterogeneous terrain make it more difficult to disentangle these technical effects from 'real' flow features, as e.g. the impact of sensor separation and flow distortion can depend strongly on wind direction. 
	
	\item \textbf{Data processing limitations:} Deriving fluxes from EC measurements involves multiple methodological choices, such as selecting an averaging period, applying a coordinate-rotation method, filtering and despiking, as well as implementing spectral and other correction schemes. In heterogeneous terrain, these choices may need to be wind-direction or flow-regime dependent and are difficult to adapt to irregular but physically meaningful events, such as drainage flows or slope winds. In particular, the choice of a suitable averaging time faces a trade-off between capturing mesoscale circulations (favoring longer averaging periods) and maintaining stationarity (favoring shorter averaging periods). 
	
	\item \textbf{Site-dependent limitations:} The EC method is applied in diverse settings (e.g. over tall canopies, in urban environments, or over water), each introducing additional limitations. For example, under nighttime stable stratification the air above a tall canopy can decouple from the within-canopy air, so that an EC measurement above the vegetation does not represent the actual surface flux \citep{Peltola2021}. In heterogeneous environments, the flux footprint varies strongly with time and flow conditions, making flux estimates highly sensitive to changes in the flux footprint extent and the respectively dominating surface types.
\end{itemize}

\subsection{Implications of EC limitations in heterogeneous terrain}
The limitations of the EC method, especially in heterogeneous terrain, manifest themselves across a wide range of applications.
Surface energy balance unclosure is a persistent issue related to EC measurements: Turbulent fluxes of sensible and latent heat typically account for only about 70-90 \% of the available energy (radiation balance minus ground heat flux), leaving a systematic residual \citep{Wilson2002,Mauder2024}. While a part of this gap stems from measurement and storage uncertainties \citep{Foken2006}, sites with larger surface heterogeneity tend to have a larger unclosure \citep{Stoy2013}. This imbalance has been linked to the omission of dispersive fluxes \citep{Margairaz2020b,Margairaz2020a}, horizontal advection, turbulent organized structures \citep{Kanda2004} and thermally-induced mesoscale circulations \citep{Wanner2022,Zhang2024}, which systematically divert parts of the energy into non-local transport pathways not captured by EC flux estimates \citep[e.g.][]{Eder2015}. 
Extending the averaging time to (multi-)day scales to capture large eddies can reduce the surface energy imbalance, but violates the stationarity assumption required by the EC method \citep{Charuchittipan2014}. \\
Under strongly stable conditions, common during calm clear-sky nights, stratification suppresses mixing and turbulence becomes intermittent and anisotropic \citep[e.g.][]{Stiperski2018,Ren2023}. Submeso-scale motions, such as gravity waves and density currents, can emerge and lead to complex wave-turbulence interactions  \citep[e.g.][]{Mahrt2014}. The coexistence of submeso-scale motions and small-scale turbulence blurs out the scale separation between mean flow and turbulence \citep[e.g.][]{Vickers2003}, making it difficult to find a suitable averaging time. The net ecosystem exchange (NEE) of CO$_2$ under stable stratification is often underestimated \citep[e.g.][]{Gu2005}, and if these biased nighttime fluxes are used to calibrate a respiration-temperature model for partitioning NEE to infer the daytime gross primary production (GPP), GPP is systematically overestimated \citep{Reichstein2005,Stoy2006}. \\ 
As the derived flux magnitudes in heterogeneous terrain depend heavily on the flux footprint \citep{Goeckede2004,Giannico2018,Tuovinen2019} -- whose size and shape are determined by time-varying meteorological conditions and surface roughness \citep{Kormann2001,Kljun2015} -- single tower measurements are less representative, complicating the comparison with coarser weather or climate models \citep{Xu2020,Chu2026}.
In these models, turbulent fluxes are subgrid-scale processes and therefore must be parameterized, typically using Monin-Obukhov similarity theory \citep[MOST,][]{Monin1954} in form of flux--variance or flux--profile relations. MOST assumes stationary conditions over homogeneous flat terrain, and relies on stability correction functions that are usually fitted to EC observations \citep[e.g.][]{Beljaars1991,Cuxart2006}. Developing such parameterizations from EC data in complex terrain is thus challenging, but it also highlights potential pathways for representing surface heterogeneity by incorporating anisotropy, coherent structures, or intermittency as additional scaling parameters \citep[e.g.][]{Stiperski2018,Stiperski2022,Ren2023,Mack2024}.

\subsection{How to make EC (more) applicable in heterogeneous environments?}
Parts of the effects of surface heterogeneity on EC measurements can be mitigated by optimizing sensor placement and site setup. For example, in complex terrain with short vegetation, EC measurements can be made very close to the surface so that the sensors remain within a shallow equilibrium layer where locally produced turbulence dominates and fluxes directly reflect the underlying surface conditions \citep{Hammerle2007,Hiller2008}. However, low measurement heights also shrink the flux footprint and make the measurement less spatially representative. To design optimal tower placement and setup, it can be helpful to perform large-eddy simulations (LES) with virtual EC towers \citep{Metzger2021,Pallandt2024}, and to deploy mobile or spatially distributed measurement systems, including auxiliary observations (e.g. for storage terms). \\
At the same time, the processing methods used to derive fluxes from high-frequency EC measurements need to be carefully tailored to the characteristics of each site \citep{Mauder2006,Goeckede2008}. Standard processing pipelines based on fixed 30 minutes temporal averaging may not be sufficient in heterogeneous terrain. Instead, (post-)processing and analysis steps of turbulence characteristics should be closely linked and iteratively refined, so that raw-data processing methods can be adapted to the specific site and environmental conditions. In this context, it is desirable that both processing and analysis routines are available as open-source software. Several packages already exist for selected applications of EC data analysis, such as \texttt{REddyProc} for gap-filling and flux partitioning \citep{Wutzler2018} and \texttt{bigleaf} for deriving ecosystem-scale physiological properties from EC data \citep{Knauer2018}. \\
Here, we build on these recent advancements by introducing the open-source software package \texttt{Reddy}. The goal is to develop a tool that combines various analyses that can be used to evaluate EC measurements and select appropriate post-processing settings, particularly in heterogeneous landscapes. Combining these functions into a single modular tool enables quick real-time application directly in the field, thereby assisting with station management from the outset and, beyond that, facilitating detailed scientific investigations. \\
We first describe the package structure and the main analysis steps, and subsequently demonstrate selected aspects of the package functionality using dual-site measurements in the alpine valley Finse (Southern Norway), a measurement campaign at the boreal lake Langtjern (Southern Norway), and observations from the permafrost peatland I\v{s}koras (Northern Norway).

\section{Package Description}
\begin{figure}
	\includegraphics[width=12cm]{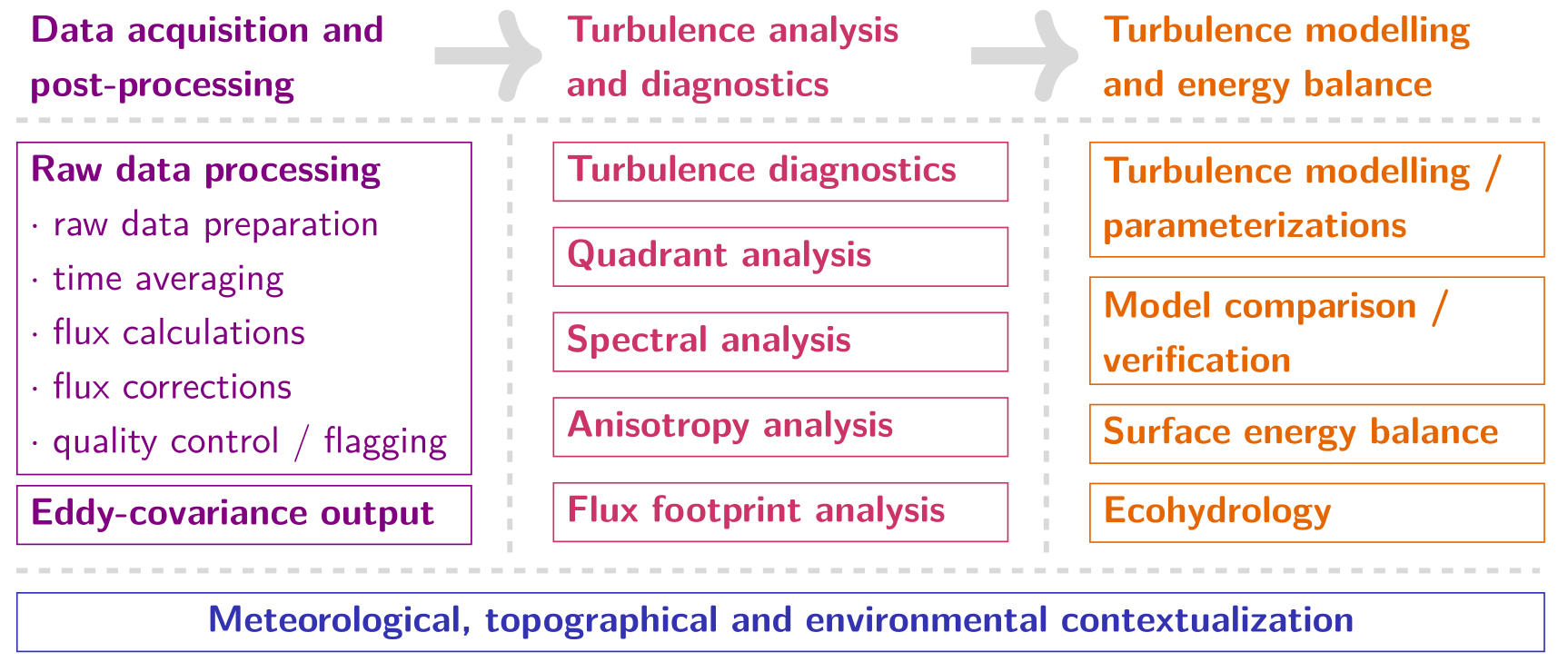}
	\centering
	\caption{Components and workflow from EC raw data to application in \texttt{Reddy}.}
	\label{fig:schema}
\end{figure}

The \texttt{Reddy} package provides functions for both raw data processing and standardized analysis of the post-processed EC data, as showcased in Fig. \ref{fig:schema} and described in the following.

\subsection{Data processing}
The raw data processing of the high-frequency EC data in \texttt{Reddy} follows the usual four-step procedure \citep[e.g.][]{Fratini2014}:
\begin{itemize} 
	\item \textbf{\underline{step1:} raw data preparation}: The raw data is despiked by using pre-defined thresholds (plausibility limits), skewness and kurtosis thresholds, and the median deviation test \citep{Mauder2013}. The used despiking thresholds can be specified for different wind sectors, stability regimes, ranges of friction velocity or divided into day- versus nighttime to handle different flow regimes. Additionally, variable conversions, e.g. from speed of sound to sonic temperature, and unit conversions are performed. The separation between gas analyzer and ultra-sonic can be corrected for with a lag-time correction applying the maximum cross-correlation method. 
	\item \textbf{\underline{step2:} tilt correction, time averaging and flux calculation}: Tilt correction is performed by rotating the axes of the three-dimensional wind measurements from the instrument frame into a coordinate system that matches the local mean flow and surface -- in heterogeneous terrain, it is thereby particularly important to define a physically meaningful vertical direction. \texttt{Reddy} provides two methods for this: (1) double rotation, to rotate the wind into a streamline-following natural coordinate system per chosen averaging time window, and (2) planar fit rotation, to rotate the wind into a mean-streamline plane \citep{Wilczak2001}. Double-rotation is recommended for flat and homogeneous terrain as well for sites with changing surface properties, e.g. during snow melt-out or growing season. Planar fit is the preferred method in complex sloped terrain where persistent 'real' vertical motions, e.g. resulting from secondary circulations and slope flows, can exist, which would be forced to zero by applying double rotation. If a site is asymmetric with different footprint characteristics, a sector-wise planar fit rotation should be considered. For heterogeneous terrain, one should systematically compare different rotation methods, including their fitted tilt angles, and then choose the configuration that yields physically consistent mean flows and fluxes across wind sectors and stability regimes. \\
	Afterwards, the means, variances and covariances of the desired quantities are calculated using block or rolling average and the covariances are transformed to fluxes. To accelerate these calculations, \texttt{Rcpp} is used which facilitates a seamless integration of R and C++ \citep{Eddelbuettel2013}. The turbulent flux calculation requires choosing an averaging time that separates turbulent from (sub)mesoscale motions. This choice is site- and flow-regime dependent \citep[e.g.][]{Moncrieff2005}, and should be based on a systematic comparison and re-evaluation of the chosen processing options, as outlined in sec. \ref{sec:re-evaluation-spiral}.
	\item \textbf{\underline{step3:} flux corrections}: Depending on the used instruments and conditions, the fluxes require specific corrections. To convert from buoyancy flux (based on sonic temperature) to sensible heat flux, SND correction \citep[named after][]{Schotanus1983} is applied, which usually includes the cross-wind correction \citep{Liu2001}.
	For the gas fluxes, volume-related quantities have to be converted to mass-related quantities by applying a density correction, which is achieved with the WPL correction (usually for open-path systems) \citep[named after][]{Webb1980} or with the adjustments of \citet{Ibrom2007} to account for de-synchronized dilution of water vapor and CO$_2$ flux (usually used for closed-path systems). 
	The limited frequency response of the measurement system acts as low-pass filter and resulting spectral loss needs to be corrected for \citep{Massmann2000}. However, when a lag-time correction with maximum cross-correlation is applied (in step 1) it corrects for both sensor separation and limited frequency response, resulting in an underestimation of high-frequency contributions, which can be accounted for with a transfer function \citep{Peltola2021b}.
	\item \textbf{\underline{step4:} quality control and flagging}: The quality control  follows the flagging system introduced by \citet{Mauder2013} applying a distortion flag, a vertical velocity flag, a stationarity flag \citep{Foken1996} and an integral turbulence characteristics flag that tests the agreement with similarity scalings from \citet{Panofsky1984}. In this 0-1-2 system, the flags take the values of 0 (good data quality), 1 (acceptable data quality) and 2 (recommended to discard). Although there is currently no dedicated flag for mesoscale secondary circulations, they often manifest through multiple of the above flags taking the value 2, for example persistent mean vertical motions and pronounced non‑stationarity. Additional quality control diagnostics provided are number of spikes and amplitude resolution.
\end{itemize}

In \texttt{Reddy}, all the described methods are implemented as separate functions, which are ultimately combined in one processing routine for 'real' EC measurements and one for virtual measurements from turbulence-resolving simulations. Due to its modular design, users can also integrate desired functions into their own customized workflows. The advantage thereby is that all processing choices remain explicit and under user control, rather than being imposed by fixed package defaults. 

\subsection{Calculation of turbulence diagnostics} \label{sec:diagnostics}

\paragraph{Turbulence diagnostics}
From the post-processed data, several standard turbulence diagnostics, such as friction velocity, turbulent kinetic energy, turbulence intensity, directional shear angle, Obukhov length, stability parameter and flux Richardson number can be calculated with \texttt{Reddy}. Over flat terrain, these diagnostics are typically interpreted as representative of the entire surface layer. In heterogeneous terrain, however, they should be computed separately for distinct stability regimes and treated as \textit{local} measures of stability. Where multi-level data are available, they should also be evaluated as height-resolved profiles to reveal possible flux divergence or flow separation. From flux-gradient relations, eddy diffusivities can be estimated by $K_x = - \overline{x'w'}/(\partial x / \partial z)$, and used to detect counter-gradient, non-local or asymmetric transport, thereby helping to assess whether the flow is consistent with (local) similarity theory or not. 
Further bulk quantities can be consulted as diagnostics in \texttt{Reddy} -- for flows under strong stratification relevant are for example: The Prandtl number ($Pr$), defined as ratio of eddy viscosity (momentum) and eddy conductivity (heat), indicates the relative efficiency of momentum and heat transport; values $Pr>1$, imply more efficient momentum than heat transport, which is typically observed under stable stratification where gravity waves enhance momentum mixing \citep{Foken2023}. The Ozmidov scale ($L_{Oz}$), defined as ratio between TKE dissipation and the cubed Brunt-Väisälä frequency, provides an estimate for the largest eddies that can still overturn against stable stratification. Since the degree of stratification controls the strength of land-atmosphere coupling \citep{Acevedo2016}, \citet{Peltola2021} introduced the decoupling metric $\Omega = \sigma_w/(\sqrt{2}zN)$, which describes whether turbulent fluctuations ($\sigma_w$) can counteract static stability described by the Brunt-Väisälä frequency ($N$). \citet{Huss2024} showed that $\Omega$ is superior to other stability metrics under strong stratification as it is in contrast to e.g. the eddy diffusivities based on the turbulence intensity and not on the fluxes. \citet{Mack2026} used this metric to relate the process of physical (de)coupling to information (de)coupling in terms of information theory, which is particularly suitable for verifying whether parameterizations or models use the available information given a physical process optimally or not.

\paragraph{Statistical flow properties}
Additionally, statistical properties of the turbulent fluctuations can be used to diagnose patchiness, intermittency, and flow persistence. For example, the skewness (3. central moment) of turbulence fluctuations (e.g. $w', T'$) indicates asymmetry in the distribution, typically associated with bursts, ejections, or thermal plumes \citep{Tennekes1972,Tillman1972}. The kurtosis (4. central moment) reflects the heaviness of the tails, indicating rare but intense events. From the autocorrelation function of turbulence fluctuations, a decorrelation time scale can be derived that characterizes the ''memory'' of the flow \citep{Agarwal2021}. Under well-developed, continuous turbulence this time scale is typically short, whereas in stable and terrain-influenced conditions it can become large, indicating persistent, organized turbulent patches or wave-like motions. \texttt{Reddy} also implements the flux-intermittency indicator from \citet{Mahrt1998}, obtained by subdividing a standard averaging period into several shorter sub-intervals thereby quantifying intermittency. 
The turbulence topology can be investigated by calculating the velocity aspect ratio \citep[e.g.][]{Mahrt2011} or by performing an invariant analysis of the full Reynolds stress tensor, which also takes the shear stresses into account (sec. \ref{kap:Reddy-analysis-methods}).

\subsection{Data analysis and visualization} \label{kap:Reddy-analysis-methods}
The quality-controlled data can be used for several applications, both based on the high-frequency data and the processed output. 

\paragraph{Spectral analysis}
\texttt{Reddy} provides functions to calculate and visualize spectra from  high-frequency measurements and numerical model output. 
Using Fast Fourier Transform (FFT), both turbulent velocity spectra (variance distribution over frequency) and co-spectra  (covariance distribution over frequency) can be analyzed, for example, to assess whether kinetic energy dissipation follows the theoretically expected slopes \citep[e.g. -5/3 in the inertial subrange,][]{Kraichnan1971}, which scales dominate the flux, and whether spectral corrections applied in the processing step need refinement. \\
While FFT assumes stationarity over the entire recorded timeseries, wavelets based on localized basis functions retain both scale and time information and are thus particularly helpful for diagnosing non-stationarity and intermittency. Multiresolution decomposition (MRD), based on Haar wavelets, is a particular popular tool to analyze EC measurements \citep{Vickers2003}. Multiple MRDs can be combined into a composite MRD, based on which a suitable averaging time for calculating fluxes and turbulence intensities can be derived \citep[e.g.][]{Stiperski2018}. Based on single and composite MRDs, \texttt{Reddy} can recommend a suitable averaging time using the algorithm of \citet{Vickers2003}, returning the first zero‑crossing (from small scales) after an identified turbulence peak, corresponding to the co‑spectral gap separating turbulent and submeso‑scale motions (example in sec. \ref{sec:langtjern}). \\
Ogives, i.e. cumulative distribution functions, offer another way to analyze spectral contributions. Although they do not contain different information than the spectra, they appear smoother due to the integration and allow an objective numerical determination of the the averaging time based on the convergence rate \citep[][]{Sievers2015}: If the ogive converges, low-frequency contributions are weak and the averaging time sufficient, while a lack of ogive convergence indicates substantial low-frequency contributions, requiring to reassess the chosen averaging time. Using ogive optimization for flux estimation is particularly suitable under low-flux conditions, when positive and negative contributions occur within the same averaging period \citep[e.g.][]{Pirk2017}. \\
Spectra can also be computed from model data, and due to the thereby available spatial information also wavenumber spectra can be calculated. For this, \texttt{Reddy} adapts the commonly used discrete cosine transform (DCT) \citep{Ahmed1974,Denis2002}. From these model spectra it is possible to assess whether the transition from large scale to mesoscale flows \citep[i.e. the transition from -3 slope to -5/3 slope,][]{Nastrom1985,Lindborg1999} is represented by the model (for both kinetic and potential energy), to determine the effective model resolution \citep[i.e. the scale where the spectrum starts to deviate from the expected slope,][]{Skamarock2004}, and thus to diagnose whether a model is overly diffusive \citep{Ricard2013,Mack2025}.



%

\paragraph{Quadrant analysis} Quadrant analysis decomposes the high-frequency data used for the flux calculation (i.e. two scalars, such as vertical velocity $w$ and a scalar $T, q, c, ...$) into four event types, allowing to assess how and by what type of structures the flux is transported \citep[e.g.][]{Thomas2007,Chowdhuri2018}. These event types relate to coherent structures, e.g. ejections, sweeps or thermal plumes, and with quadrant analysis their intensity and occurrence frequency can be systematically studied. \texttt{Reddy} provides the option to visualize a normalized scatter plot indicating the number and flux contributions of the four quadrants for different hyperbolic hole size (i.e. intensities) and to derive simple measures for flow organization, such as ejection-sweep ratio, exuberance \citep{Shaw1983} or organization ratio \citep{Mack2024}. In heterogeneous environments where source-sink distributions for different scalars are not congruent (e.g. cities, or distinct emission hotspots of a gas in a thermally heterogeneous environment), the scalars can exhibit different transfer efficiencies, which can be examined in detail using quadrant analysis \citep{Schmutz2019,Mack2024}.

\paragraph{Structure functions} Structure functions of different orders can be calculated \citep[e.g.][]{vandeWater1999} as well as two-point correlations \citep[e.g.][]{Ganapathisubramani2005}. 
A structure function quantifies how differences in a scalar field grow with distance. For a scalar $c$, the second-order structure function at a separation distance $r$ is defined as $D_{cc}(r) = \langle (c(x+r)-c(x))^2\rangle$. In the inertial subrange it follows a 2/3-law, expressed via the structure parameter $C_c^2=D_{cc}(r)r^{-2/3}$. From structure functions, the degree of heterogeneity of different variables can be assessed (large $D_{cc}$ = strong spatial variability, small $D_{cc}$ =  close to homogeneity), and the agreement with the theoretical inertial-subrange scaling can be tested. In contrast to structure functions, which quantify the magnitude of differences, two-point correlations, i.e. $R_{cc}=\mathrm{cor}(c(x),c(x+r))$, quantify the degree of similarity or coherence of the $c$-field between locations separated by $r$.

\paragraph{Invariant analysis of the Reynolds stress tensor, anisotropy and barycentric map}
From an invariant analysis of the Reynolds stress tensor $R=\overline{u_i'u_j'}$ derived eigenvalues and eigenvectors characterize turbulence geometry and spatial orientation of turbulent eddies. In a barycentric map \citep{Banerjee2007}, different anisotropy limiting states corresponding to distinct turbulence geometries can be visualized (3‑component: isotropic turbulence, 2‑component: pancake‑like axis-symmetric turbulence, 1‑component: wave‑like motions). Anisotropy quantifies deviations from idealized isotropic (spherical) turbulence, and incorporating the occurrence frequency of the different anisotropy limiting states has been shown to improve MOST scalings in both homogeneous \citep{Stiperski2018} and heterogeneous terrain \citep{Stiperski2022}. In shear‑dominated flows, 2‑component eddies tend to dominate, whereas under strong stratification wave‑like motions lead to 1‑component turbulence \citep{Vercauteren2019}; combined with spectral analysis, this can be used to diagnose the origin of low‑frequency submeso‑scale flux contributions \citep{Mack2024}.

\paragraph{Flux footprint}
The flux footprint, i.e. the area where the measured flux originates from, can be estimated based on the processed EC data. In \texttt{Reddy}, the analytical flux footprint model from \citet{Kormann2001} with the parameters according to \citet{Neftel2008}, and the parameterized version of a Lagrangian stochastic model from \citet{Kljun2015} are available. For both, the cross wind-integrated flux footprint (1D flux footprint) and the contours (2D flux footprint) can be calculated, and afterwards geo-localized to be plotted on terrain maps. For a time series of processed EC data, a flux footprint climatology can be calculated for both footprint models. These flux footprint parameterizations are derived for stationary conditions and for homogeneous surfaces, and can only incorporate a displacement height. However, in combination with land cover maps, the flux footprint can be disaggregated into different fluxes per surface type using data-driven approaches, thus accounting for land surface heterogeneity within a flux footprint \citep{Suehring2019,Pirk2024,Schlutow2026}.

\paragraph{Surface energy balance} If additional to the flux measurements also radiometer (and ground heat flux) measurements are available, the surface energy balance can be studied by calculating and visualizing the closure ratio and the residual flux. The residual flux is usually larger at heterogeneous than homogeneous sites \citep[][]{Stoy2013}, which can be traced back to the measurement setup, i.e. the EC measurements capture a different footprint than the radiometer or ground heat flux measurements \citep{Mauder2006,Mauder2020}, and to the omission of non-local transport, such as horizontal advection and mesoscale organized structures \citep{Eder2015,Wanner2022,Mauder2024}. The closure ratio can thus also be used as diagnostics for the degree of surface heterogeneity \citep[e.g.][]{Hammerle2007}.

\paragraph{Model evaluation tools and similarity scalings} EC data is widely used to develop turbulence parameterizations for NWP or climate models. Accordingly, \texttt{Reddy} provides basic tools to analyze MOST similarity relations for both flux-variance and flux-profile relations, calculate dimensionless shear and temperature gradients, and derive theoretical vertical wind and eddy diffusivity profiles. As MOST-based similarity scalings assume stationary and horizontally homogeneous conditions, EC observations in combination with measured bulk variables (e.g. slowly sampled temperature and wind speed measurements at different heights) are particularly suitable for studying deviations from these idealized scalings -- examples for this are including anisotropy \citep{Stiperski2022} or a measure for organization of coherent structures \citep{Mack2024} as additional scaling parameter (alongside the stability parameter) to account for terrain-modified turbulence geometry or intermittent bursting. 
\texttt{Reddy} further provides functions to process virtual measurements from turbulence-resolving simulations (e.g. LES), enabling studying spatial flux variability or optimizing flux tower placement \citep[e.g.][]{Xu2020}. 
For comparing NWP model output with observations, standard functions for e.g. the conversion of the used vertical model coordinate and the deaccumulation of fluxes \citep{Mack2025} are available within \texttt{Reddy}.

\subsection{Environmental controls}

\paragraph{Topographic indices}
\texttt{Reddy} can compute several topographic indices from digital elevation models (DEM) that describe the position of each point relative to its surroundings (e.g. ridge, slope, or valley). These indices are useful for characterizing the potential for cold-air pooling, drainage pathways and terrain-controlled flows. Examples of metrics include the topographic position index (TPI), the deviation from mean elevation (DEV), and the slope-based direction-dependent terrain parameter Sx \citep[a shelter index introduced by][]{Winstral2017}. In addition, surface roughness length can be estimated from friction velocity using the relation of \citet{Charnock1955}. 

\paragraph{Meteorological and ecohydrological contextualization}
Apart from regime classifications, further standard quantities based on auxiliary measurements can be consulted for a meteorological contextualization of the EC measurements, e.g., clear-sky index \citep[CSI,][]{Marty2000,Lehner2019}, vapor pressure deficit (VPD), saturation vapor pressure and several unit conversions, e.g. of different humidity measures, can be performed. 
For a hydrological contextualization, the calculated latent heat flux can be transformed to evaporation (or evapotranspiration ET) and in combination with the sensible heat flux, Bowen ratio and evaporative fraction can be calculated. 
GPP can be partitioned from CO$_2$ fluxes in two steps \citep{Reichstein2005}: (1) Using nighttime measurements (when photosynthesis is negligible, i.e. NEE equals the ecosystem respiration Reco) to fit a temperature--respiration relation, and (2) fitting a daytime light-response curve of NEE versus PAR (photosynthetically active radiation). Combining these, Reco can be predicted for each daytime period and GPP can be obtained as GPP $=$ Reco -- NEE.
At heterogeneous sites, these ecohydrological relations should be subdivided by flow regimes and flux footprint composition. For example, GPP--VPD or GPP--PAR relations can be fitted separately for periods when the footprint is dominated by different land-cover classes, so that the parameterizations reflect the behavior of each dominant surface type.

\subsection{Combining processing and analysis steps in a re-evaluation spiral} \label{sec:re-evaluation-spiral}
\begin{figure}
	\includegraphics[width=10cm]{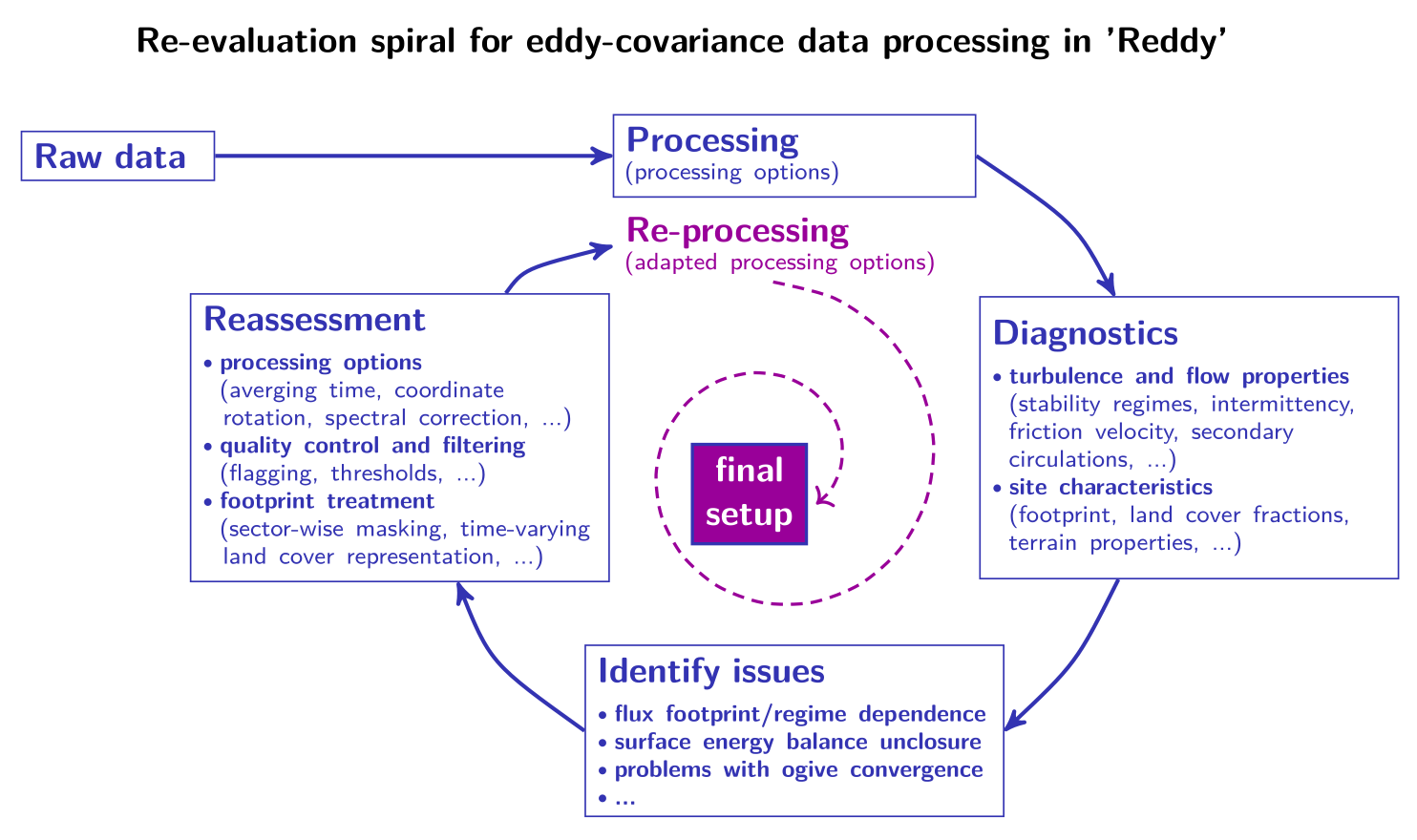}
	\centering
	\caption{Re-evaluation spiral for the post-processing of EC measurements in \texttt{Reddy}.}
	\label{fig:reevaluation-spiral}
\end{figure}

Compared to existing software packages, \texttt{Reddy} is built in a modular way and unifies both post-processing and analysis steps, enabling users to tailor the processing choices to the specific site and meteorological conditions. Conceptually, this workflow follows a re-evaluation spiral (Fig. \ref{fig:reevaluation-spiral}). First, the raw data is processed using a standard (default) configuration. From these preliminary fluxes, turbulence diagnostics are computed and standard analyses are visualized. This initial analysis step is used to identify potential issues and site characteristics, such as regime and flux footprint dependence of the fluxes, surface energy balance (un)closure or problems with ogive convergence. Based on these insights, the original processing configuration is revised, and key choices, e.g. the averaging time, the rotation method, or the subdivision into different wind sectors, are progressively refined. Iterating this loop leads to a site- and flow-tailored processing setup.

\subsection{Manual and Documentation}
The \texttt{Reddy}-package is in line with the standards from R-CRAN (Comprehensive R Archive Network) and utilizes the \texttt{roxygen2}-package for standardized in-code comments to create a detailed PDF manual. This manual contains a detailed description of all functions, their inputs and outputs, accompanied by examples and automatized testing procedures. Additionally, a \texttt{bookdown}-generated GitHub-page (\url{https://noctiluc3nt.github.io/ec_analyze/}) showcases and explains the \texttt{Reddy}-package grouped by topics (detailed in sec. \ref{sec:notebooks-teaching}).

\section{Applications}
\begin{figure}
	\centering
	\includegraphics[width=10cm]{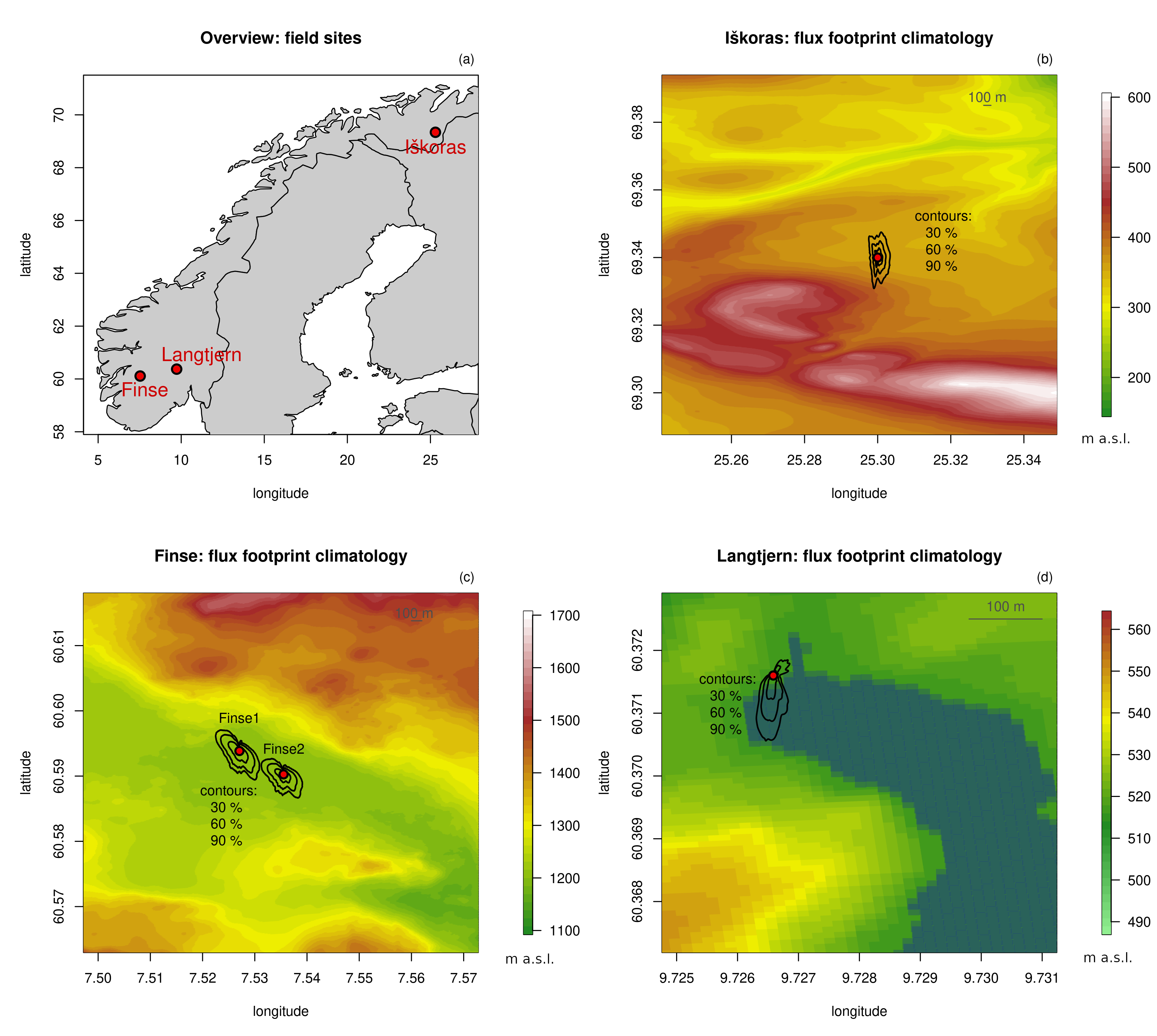}
	\caption{Overview of the studied field sites and their flux footprint climatologies with respective topography: (a) Site locations, (b) I\v{s}koras (used time period: 01/2019-12/2020), (c) Finse (used time period: 02/2018-01/2019), (d) Langtjern (during the measurement campaign). Different scales.}
	\label{fig:sites}
\end{figure}

We showcase parts of the \texttt{Reddy}-package based on EC measurements from three different locations in Norway (Fig. \ref{fig:sites}a). Finse (60.11$^\circ$N, 7.53$^\circ$E, 1200 m a.s.l.) is an alpine tundra site on the Hardangervidda mountain plateau in Southern Norway. Here, we use measurements from two neighboring sites (distance of 610 m) both equipped with an ultra-sonic (CSAT3, Campbell Scientific) and a closed-path gas analyzer for H$_2$O and CO$_2$ (Li-7200, LiCor), as well as slow-sampled auxiliary measurements described in \citet{Pirk2023} and \citet{Mack2024}. We investigate a two-day case study during clear-sky conditions (17.03.-18.03.2018) focusing on using \texttt{Reddy} for calculating turbulence diagnostics and surface energy balance closure. \\
Additionally, we use data from a measurement campaign (12.04.2024-24.06.2024) with an ultra-sonic (CSAT3, Campbell Scientific) and an open-path gas analyzer for H$_2$O and CO$_2$ (Li-7500, LiCor) at lake Langtjern (60.37$^\circ$N, 9.73$^\circ$E, 517 m a.s.l.) in Southern Norway. This lake is home to a water and climate monitoring station studying the exchange of terrestrial carbon at this lake--forest--atmosphere interface \citep[e.g.][]{Clayer2021,deWit2024}. While carbon concentration measurements are routinely performed along a vertical profile through the lake, EC measurements are usually not available. The measurement period encompasses the transition between ice-covered and ice-free conditions and we use \texttt{Reddy} to investigate turbulence spectra before and after ice-cover transition. \\
Lastly, we use two years (2019-2020) of continuous measurements with an ultra-sonic (CSAT3, Campbell Scientific) and two gas analyzers (closed-path Li-7200 for H$_2$O and CO$_2$, and Li-7700 for CH$_4$, LiCor) from I\v{s}koras (69.34$^\circ$N, 25.30$^\circ$E, 380 m a.s.l), a permafrost peatland site in Northern Norway \citep{Pirk2024}, to analyze flux-variance relations of trace gases with the help of \texttt{Reddy}. \\
We apply \texttt{Reddy}'s re-evaluation spiral to determine the processing options set in \texttt{ECprocessing()}, which applies despiking (\texttt{despiking()}), double rotation (\texttt{rotate\_double()}) or planar-fit rotation \linebreak(\texttt{rotate\_planar}), SND and WPL correction (\texttt{SNDcorrection()}, \texttt{WPLcorrection()}), quality flagging (e.g. \texttt{flag\_stationarity()}, \texttt{flag\_distortion()}), necessary variable and unit conversions, cross-correlation maximization (\texttt{shift2maxccf()}) and calculations of turbulence diagnostics. 
The flux footprint climatologies shown in Fig. \ref{fig:sites} are calculated with the function \texttt{calc\_flux\_footprint\_climatology}\linebreak\texttt{(..., method = "KM2001")}, here exemplary with the \citet{Kormann2001}-model, and geo-localized (\texttt{locate\_flux\_footprint()}), and indicate valley-following flow channeling at Finse and a lake--dominated fetch at Langtjern.

\subsection{Example 1: Finse} \label{sec:finse}
\begin{figure}
	\centering
	\includegraphics[width=10cm]{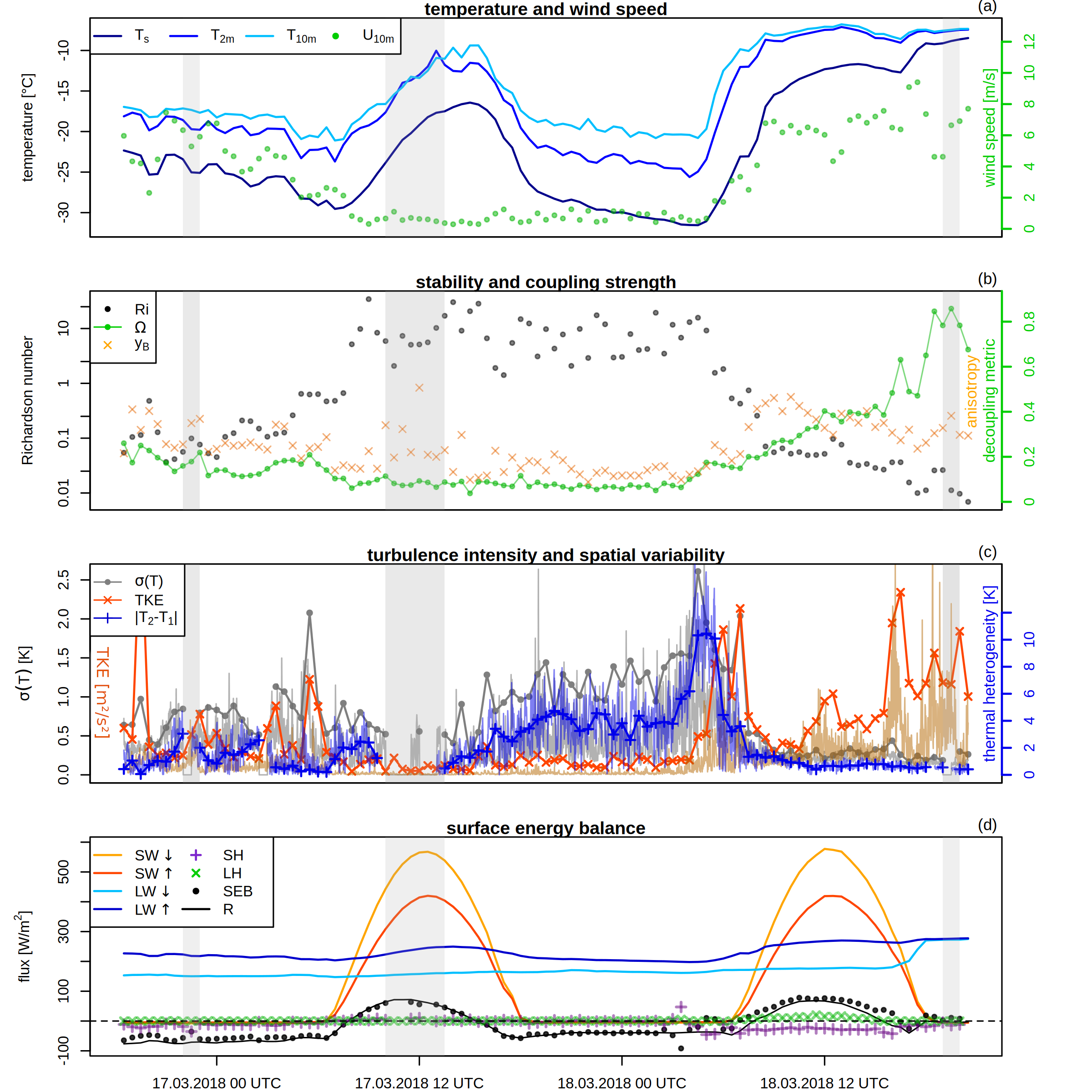}
	\caption{Case study from Finse, 17.03.2018-18.03.2018:  (a) Temperatures ($T_s$: surface temperature, $T_{2\,\textrm{m}}$: 2 m-air temperature, $T_{10\,\textrm{m}}$: 10 m-air temperature) and wind speed ($U_{10\,\textrm{m}}$), (b) Richardson number and decoupling metric (based on vertical gradients between atmosphere (10 m) and surface) and anisotropy ($y_B$, from invariant analysis of the Reynolds stress tensor), (c) turbulence intensities and 2 m-air temperature difference between the two sites Finse1 ($T_1$) and Finse2 ($T_2$) (30 minutes and 1 minute temporal averaging), (d) surface energy balance. Shaded areas mark bad quality flags (QF = 2).}
	\label{fig:finse}
\end{figure}

Fig. \ref{fig:finse} shows a two-day time series generated with \texttt{Reddy} for very cold conditions at Finse during a high-pressure episode over Scandinavia (17.03.-18.03.2018) -- for investigating \textit{how turbulence diagnostics evolve during a clear-sky stably stratified night in the alpine valley Finse}.
From the processed EC data several turbulence diagnostics (details Tab. \ref{tab:diags-app}) and the surface energy balance were calculated and visualized (\texttt{calc\_seb()}, \texttt{plot\_seb()}). These two clear-sky days exhibit a pronounced diurnal cycle and very stable stratification during the calm night, with Richardson numbers exceeding $Ri>10$ (Fig. \ref{fig:finse}b). While during daytime the atmosphere is well-mixed (similar air temperatures at 2 m and 10 m, Fig. \ref{fig:finse}a), surface and atmosphere remain largely decoupled throughout the entire day (Fig. \ref{fig:finse}a, b). This is reflected by low values of the decoupling metric $\Omega$ indicating atmosphere-surface decoupling, related to weak turbulence (low TKE values, Fig. \ref{fig:finse}c) that is insufficient to act against the strong vertical temperature gradient. At the same time, the turbulence is strongly anisotropic (low values of $y_B$, Fig. \ref{fig:finse}b). Temperature fluctuations ($\sigma(T)$, and thus turbulent potential energy, Fig. \ref{fig:finse}c) are largest during the low-wind and weak turbulence period. The 2 m-air temperature difference between the two sites (Finse1 and Finse2, Fig. \ref{fig:finse}c) is small during daytime, but increases markedly at night, reaching a maximum of up to 10 K (over a horizontal distance of 610 m). At the onset of the morning transition (18.03.2018 06 UTC), TKE rises rapidly, atmospheric mixing increases, Ri drops below $Ri<0.25$, and the inter-site temperature difference reduces (to less than 1 K). Structurally, this inter-site temperature difference corresponds to a simple first-order structure function and thus provides a measure of thermal heterogeneity. 
The surface energy balance (Fig. \ref{fig:finse}d) is dominated by radiative fluxes, with radiative cooling during nighttime and heating during daytime, likely contributing to snow-melt. The turbulent fluxes remain small (amplitudes smaller than 20 W/m$^2$) and the ground heat flux is negligible ($<$ 2 W/m$^2$, not shown), so the surface energy balance cannot be closed. For these two days, the mean residual flux is $-$11.3 W/m$^2$, corresponding to an unclosure of 10.8 \%. \\
Overall, during this clear-sky, stably stratified night at Finse, turbulence diagnostics indicate very weak, strongly anisotropic turbulence and pronounced atmosphere--surface decoupling accompanied by large horizontal temperature differences between sites, followed by a rapid morning transition with increasing TKE, renewed mixing, reduced stability and a collapse of the thermal heterogeneity.

\subsection{Example 2: Langtjern} \label{sec:langtjern}
\begin{figure}
	\centering
	\includegraphics[width=10.5cm]{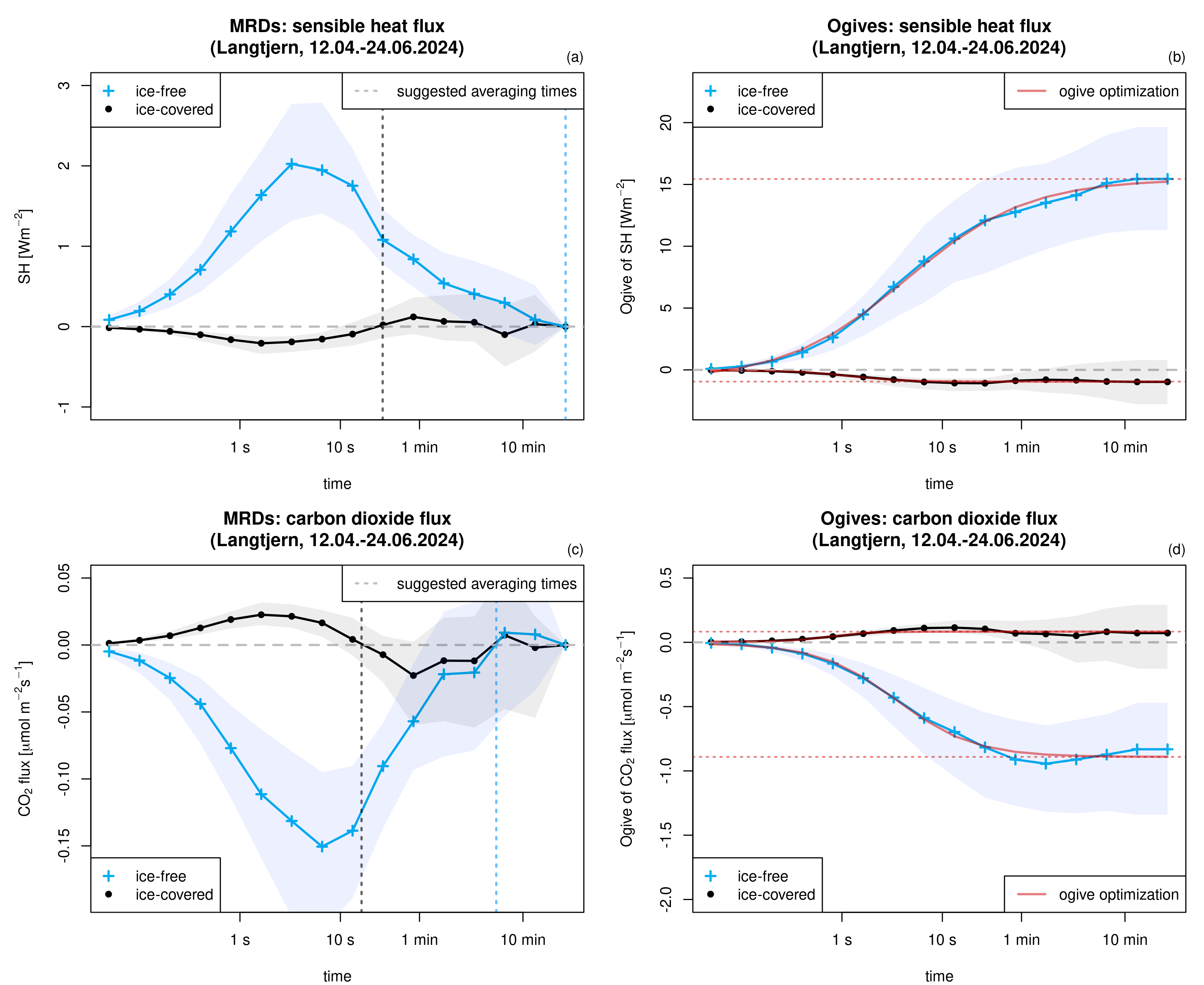}
	\caption{Composite MRDs and ogives during ice-covered and ice-free period at lake Langtjern. The shaded areas mark the inter-quartile-range and the solid vertical line the suggested averaging time.}
	\label{fig:langtjern}
\end{figure}

Here, we use \texttt{Reddy} to investigate \textit{how ice-covered versus ice-free conditions affect the relative contributions of turbulent and submeso-scale motions to sensible heat and CO$_2$ fluxes at lake Langtjern.}
For this, Fig. \ref{fig:langtjern} shows the composite MRDs (\texttt{calc\_mrd()}) and ogives (\texttt{calc\_ogive()}) of sensible heat and CO$_2$ fluxes during ice-covered (until 03.05.2026) and subsequently ice-free conditions at lake Langtjern. 
During the open-water period, the MRDs of sensible heat show predominantly upward (positive) fluxes, while the CO$_2$ flux is mostly negative. Episodes of positive CO$_2$ flux were observed immediately after the ice break-up, indicating transient outgassing followed by a spring overturning of the lake water column \citep{Clayer2021,Vogt2026}. The subsequently sustained negative CO$_2$ fluxes coincide with a cold spell in late May, where lower water temperatures increase CO$_2$ solubility, leading to a temporarily CO$_2$-undersaturation of the lake. This pattern is consistent with previous observations from an alpine lake by \citet{Scholz2021}.
During ice cover, flux magnitudes are smaller with significant submeso-scale contributions. Following \citet{Vickers2003}, the first zero-crossing of the MRDs (starting from the highest frequency) indicate a scale separation between turbulence and submeso-scale, representing a suggestion for a suitable averaging time (\texttt{suggest\_avgtime\_from\_mrd()}). With this approach, averaging times of about 30 s are suggested for both fluxes during the ice-covered period. At these turbulent scales ($<$ 30 s), the sensible heat is negative (stable stratification), while the CO$_2$ flux is positive (CO$_2$ release). At longer (submeso) scales, both fluxes exhibit contributions of the opposite sign. These submeso motions are excluded when using a small averaging time of 30 s but included in the turbulent flux when using a standard averaging time of 30 minutes, leading to a partial cancellation of turbulent and submeso contribution. 
Strong non-local, low-frequency contributions to air--lake CO$_2$ fluxes have been reported previously \citep[e.g.][]{Esters2020,Scholz2021} and linked to the sharp forest-lake transition and a short fetch, so that air parcels have a short "time over water" before being sampled. Such effects are likely important at Langtjern, where the opposite shore is less than 300 m from the flux tower. An additional indicator of non-local effects is the skewness of high-frequency specific humidity ($S_q$, sec. \ref{sec:diagnostics}) \citep{vandeBoer2014}. Negative $S_q$ (a dry tail) indicates entrainment of dry air into the internal lake boundary layer or horizontal advection of drier air from the land over the lake, whereas near-zero or positive skewness are consistent with locally generated surface-layer turbulence. Similar to \citet{Esters2020}, we find predominantly negative $S_q$ during the ice-covered period, but positive $S_q$ during the ice-free period. \\
Adjusting the averaging time can mitigate some submeso contamination of the flux estimate, but cannot fully separate turbulent and submeso contributions. \citet{Sievers2015} therefore proposed using ogive optimization, which estimates fluxes directly from the ogive without making assumption about a scale gap.
For this, a theoretical sigmoidal co-spectral model is fitted to the ogives (\texttt{fit\_sigmoid\_ogive()}, Fig. \ref{fig:langtjern}b,d). The asymptote of the fitted sigmoid curve (ogive convergence, dotted red lines in Fig. \ref{fig:langtjern}b,d) is then taken as the flux estimate. 
The different flux estimation methods (1: standard 30 minutes averaging, 2: averaging time adapted based on \citet{Vickers2003}-algorithm, 3: flux estimate from ogive optimization) lead here to a difference of about 20\,\% in the flux magnitudes. With \texttt{Reddy} it is possible to apply all three methods, compare them systematically and derive the most appropriate processing option for the respective site and conditions.\\
Overall, during the ice-covered period turbulence is confined to small scales ($<30$ s), and sensible heat and CO$_2$ fluxes show strong submeso-scale contributions, consistent with a negatively skewed humidity distribution, whereas during ice-free conditions turbulence is well developed and a standard 30-minute averaging time appears appropriate.

\subsection{Example 3: I\v{s}koras} \label{sec:iskoras}
\begin{figure}
	\centering
	\includegraphics[width=10cm]{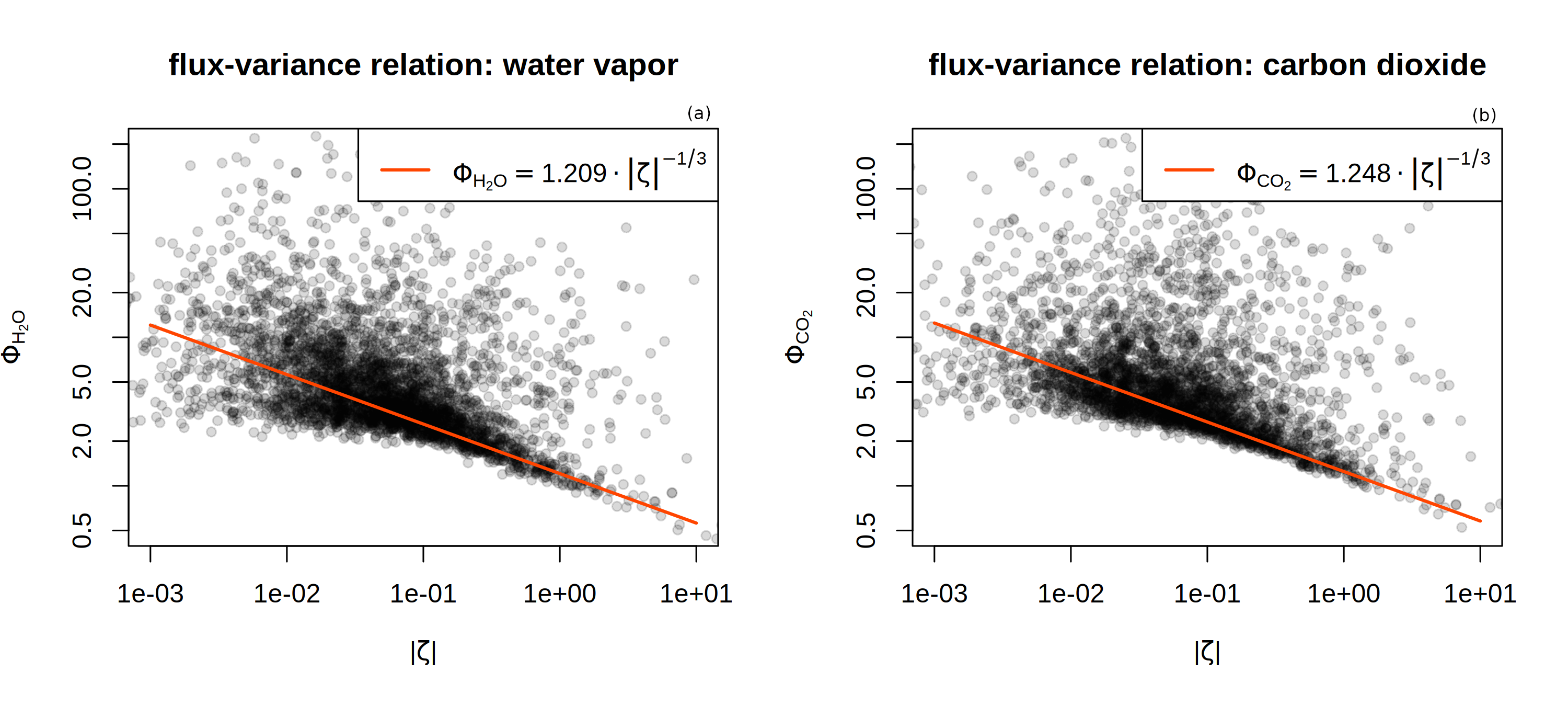}
	\caption{Flux-variance relations for (a) water vapor and (b) carbon dioxide fitted based on two years (2019-2020) of EC measurements at I\v{s}koras.}
	\label{fig:iskoras}
\end{figure}

Here, we use \texttt{Reddy} to investigate \textit{how well MOST-based flux-variance relations predict scalar fluxes and what they reveal about relative transfer efficiencies at the palsa mire I\v{s}koras} (Fig. \ref{fig:iskoras}). 
Flux-variance relations are based on MOST \citep{Tillman1972} and relate the flux of a scalar $\overline{w'x'}$ to the temporal variance of the scalar $\sigma_x$, given the stability parameter $\zeta$ and the friction velocity $u_*$ (formulas see Tab. \ref{tab:diags-app}). The structural form of $\Phi_x$ and its parameters need to be fitted to observations (at sites where the flux is measured), but if $\Phi_x$ is found to be generalizable than it can be used to estimate fluxes from single-height variance measurements. However, the degree of generalizability of $\Phi_x$ for different scalars, different sites and different flow conditions is an ongoing debate and tight to limitations of MOST \citep[e.g.][]{Katul1994,Hsieh2008,Stiperski2022}. 
Since I\v{s}koras is subject to strong seasonality in snow cover, we focus on the snow-free summer period and thus use the flux-variance relations for the convective $-1/3$-power scaling in the form $\Phi_x = C_x \cdot \vert \zeta \vert^{-1/3}$ \citep{Hsieh2008}.
From linear regression, we estimate the constant $C_x$ for $x=$ T, H$_2$O, CO$_2$, CH$_4$, as exemplified in Fig. \ref{fig:iskoras} and listed in Tab. \ref{tab:flux-variance}. The resulting $C_x$'s for I\v{s}koras are within the range of reported values in the literature, with comparably good $R^2$ \citep{Hsieh2008,Wang2012}. 
When validating the estimated fluxes using the fitted flux-variance relations compared to the measured fluxes, correlations of 0.91 for H$_2$O, 0.51 for CO$_2$ and 0.14 for CH$_4$ are achieved, indicating limits in predicting CO$_2$-fluxes and particularly CH$_4$-fluxes. This is likely related to their heterogeneous source-sink distribution: Given a homogeneous surface with the same (congruent) scalar emission fields should result in the same $C_x$ for all (passive) scalars, while different effective source areas within the same geometric flux footprint result in different transfer efficiencies and different $C_x$ for different scalars. The ratio of $C_x/C_T$ is commonly used to compare the passive scalar transfer efficiency with the heat transfer efficiency \citep[e.g.][]{Katul1994,Hsieh2008,Mack2024}, and indicates that at our palsa mire site H$_2$O is 11\,\%, CO$_2$ 14\,\% and CH$_4$ 41\,\% less efficiently transported than heat. By disaggregating the flux contributions within the flux footprint area at I\v{s}koras in ponds, palsas and mires, \citet{Pirk2024} explicitly quantified the different CO$_2$ and CH$_4$ emission rates per surface type, confirming the strong sink-source heterogeneity. \\
Overall, flux-variance relations allow accurate estimation of latent heat flux at I\v{s}koras (and to lesser extent CO$_2$ and CH$_4$ fluxes), while the heterogeneous source-sink distributions leads to different scalar transfer efficiencies for each gas.

\begin{table} 
	\centering
	\caption{Coefficients in the flux-variance relations for $T$, H$_2$O, CO$_2$ and CH$_4$ (and respective $R^2$-value) in the form $\Phi_x = C_x \cdot \vert \zeta \vert^{-1/3}$, fitted for the palsa mire site at I\v{s}koras and compared to literature values from different surface types.}
	\begin{tabular}{l|ll|ll|ll|ll|l}
		\textbf{ecosystem} & $C_T $  & ($R^2$) & $C_{\mathrm{H_2O}} $  & ($R^2$) & $C_{\mathrm{CO}_2} $ & ($R^2$)& $C_{\mathrm{CH}_4} $& ($R^2$) & \textbf{data}\\
		\hline
		palsa mire & 1.092& ($0.91$) &  1.209 &($0.85$) & 1.248 &($0.87$) & 1.540 &($0.79$) & I\v{s}koras, 2019-2020\\
		\hline
		grassland & 1.1 & ($0.64$) & 1.1 &($0.60$) & 0.95 &($0.16$)& ---& & \citet{Hsieh2008} \\
		rice field & 1.0 &($0.60$) & 1.0 &($0.78$) & 1.0 &($0.07$) & ---& & \citet{Hsieh2008} \\
		forest & 1.25 & ($0.87$) & 1.5 &($0.61$) & 1.7&($0.42$) & ---& & \citet{Hsieh2008} \\
	\end{tabular}
	\label{tab:flux-variance}
\end{table}

\subsection{Further Applications} \label{sec:notebooks-teaching}

\paragraph{Model verification, development and data assimilation}
\texttt{Reddy} also provides functions to handle model output, example applications include:
\begin{itemize}
	\item Using turbulence-resolving simulations (e.g. LES) to optimize flux tower placement before installation, or to infer source areas systematically using data assimilation techniques \citep[e.g.][]{Pirk2022}.
	\item Extracting turbulence parameterizations from NWP or climate models and evaluating them against EC data, in order to (i) identify limitations in heterogeneous or complex terrain, and (ii) implement probabilistic approaches that account for uncertainties in the representation of subgrid-scale processes \citep[e.g.][]{Mack2026}.
\end{itemize} 

\paragraph{Jupyter notebooks, in-field data processing and station monitoring}
Due to the modular structure of the \texttt{Reddy}-package, it enables several applications ranging from quick in-field real-time post-processing of measurement data to station management and detailed scientific investigations. To allow a quick introduction to the package and standard analysis procedures of EC measurements, the following Jupyter notebooks (available at \url{https://noctiluc3nt.github.io/ec_analyze/}) have been compiled covering the topics: \\

\begin{compactenum}
	\item[(1)] Raw data processing
	\item[(2)] Basic turbulence diagnostics
	\item[(3)] Quadrant analysis
	\item[(4)] Spectral analysis
	\item[(5)] Flux footprint estimation
	\item[(6)] Invariant analysis of the Reynolds stress tensor
	\item[(7)] Surface energy balance 
	\item[(8)] Turbulence parameterizations in numerical weather prediction and climate models \\
\end{compactenum}

This low-threshold introduction is aimed at new users of EC data and was used at the University of Oslo as part of teaching on EC data analysis as tool in boundary-layer meteorology in connection with a field excursion. Ready-made evaluation tools and functions enable rapid learning progress and at the same time give interested students the opportunity to engage with the source code, so that EC data analysis is not just treated as a black box, representing an advantage compared to purely GUI-based (GUI: graphical user interface) EC software (sec. \ref{sec:existing-software}).

\section{Discussion}

Because EC measurements are crucial for studying surface–atmosphere exchanges among the hydrosphere, biosphere, cryosphere, and atmosphere, a number of software packages have been developed. Below, these tools are briefly summarized (with further details available in the cited literature), and the novel aspects of \texttt{Reddy} are highlighted.

\subsection{Existing software}\label{sec:existing-software}

The following list gives a brief overview over other EC software, divided into (1) processing of the raw high-frequency measurements, (2) post-processing and analysis of the processed flux data, (3) specific applications, and (4) data portal access.

\paragraph{Raw-data processing:} The main focus of the following tools is to derive flux products from the high-frequency raw data:
\begin{compactenum}
	\item[--] \textbf{\texttt{EddyPro}} (Fortran): Processing of EC raw data provided by the manufacturer LI-COR Biosciences, including an intuitive GUI.
	\item[--] \textbf{\texttt{TK3}} (Fortran) \citep{Mauder2011TK3}: Processing of EC raw data with extended quality control tools.
	\item[--] \textbf{\texttt{EddyUH}} (Matlab) \citep{Mammarella2016}: Processing of EC raw data, including more research-focused analysis tools, requiring a Matlab license.
	
	\item[--] \textbf{\texttt{RFlux}} (R): GUI for post-processing of EC raw data by calling EddyPro.
	\item[--] \textbf{\texttt{eddy4R}}-Docker (R) \citep{Metzger2017}: Family of several R-package for EC processing in a DevOps framework.
	\item[--] \textbf{\texttt{PyFluxPro}} (Python) \citep{Isaac2017}:  Processing of EC raw data designed and developed by the OzFlux community (Ozone-focused).
\end{compactenum}

\paragraph{Post-processing:} The following software packages focus on EC data post-processing:
\begin{compactenum}
	\item[--] \textbf{\texttt{ONEFlux}} (C) (“Open Network-Enabled Flux processing pipeline”): Post-processing of (half-)hourly EC data used to create the FLUXNET2015 dataset \citep{Pastorello2020}.
	\item[--] \textbf{\texttt{REddyProc}} (R) \citep{Wutzler2018}: Post-processing of (half-)hourly EC measurements with focus on $u_*$-filtering \citep{Papale2006,Barr2013}, gap-filling \citep{Reichstein2005} and flux partitioning \citep{Reichstein2005,Lasslop2010}. 
	\item[--] \textbf{\texttt{openeddy}} (R): Post-processing and quality control of EC data \citep{McGloin2018} including visualizations, aligned with \texttt{REddyProc}.
	\item[--] \textbf{\texttt{flux-data-qaqc}} (Python) \citep{Volk2021}: Object-oriented Python package with possibility of interactive visualization for post-processing of EC measurements to derive daily or monthly evapotranspiration in an energy balance framework, which is used for generating a benchmark ET dataset within the OpenET project \citep{Melton2022}. 
	
\end{compactenum}

\paragraph{Specific Applications:} The following software focuses on specific applications using processed data:
\begin{compactenum}
	\item[--] \textbf{\texttt{bigleaf}} (R) \citep{Knauer2018}: Calculation of physiological ecosystem properties from EC measurements in a 'bigleaf'-framework, which is commonly used in land surface models to infer bulk ecosystem properties.
	\item[--] \textbf{\texttt{FluxnetLSM}} (R) \citep{Ukkola2017}: Preparing FLUXNET data for use in land surface models by e.g. gapfilling it.
\end{compactenum}

\paragraph{Data portal access:} The following software is designed for downloading flux data from online data portals.
\begin{compactenum}	
	\item[--] \textbf{\texttt{icoscp}} (Python, R): Access to EC data via the ICOS data portal.\\
\end{compactenum}

These software packages have the advantage of being extensively tested and widely used. However, many are not open‑source, require specific (meta)data formats (e.g. \texttt{openeddy}), or are not modular, which limits modifiability and flexibility. Some are strongly GUI‑oriented (e.g. \texttt{EddyUH}), while others exist only as command‑line tools (e.g. \texttt{TK3}). In addition, some of the software packages are tailored to specific applications (e.g. \texttt{FluxnetLSM}, \texttt{bigleaf}) or trace gases (e.g. \texttt{PyFluxPro}), or predefined methodological choices (e.g. \texttt{REddyProc}). For example, \texttt{REddyProc} routinely applies $u_*$-filtering to discard conditions with poorly developed turbulence, which can improve GPP estimates but may also remove measurements that are valuable for studying turbulence under stable stratification.

\subsection{\texttt{Reddy} in comparison to existing software}
\texttt{Reddy} is a fully open‑source framework for reproducible (post‑)processing and analysis of EC data. It has been successfully used in scientific studies, near‑real‑time station monitoring, and teaching.
The package is built in a modular way and offers a broad set of functions for different (post‑)processing steps -- it allows users to assemble pipelines tailored to their specific sites and research questions, and to integrate these into their own workflows.
\texttt{Reddy} brings together methods from meteorology, hydrology and ecology -- compared to existing packages, more emphasis is given to turbulent flow properties. For example, it provides tools for
turbulence diagnostics,  anisotropy and quadrant analysis, enabling to study more complex flows, e.g. lateral advective fluxes and intermittency. 
\texttt{Reddy} can also be used for processing virtual EC measurements from LES and for comparison with NWP models thus supporting model verification and development. 
At the same time, effective use of \texttt{Reddy} requires a solid understanding of EC data analysis; to support this, hands‑on Jupyter notebooks illustrating the key analysis steps are provided. The overall concept -- modular design, diverse function pool, model‑evaluation tools, and detailed documentation -- makes \texttt{Reddy} a useful extension to existing software, particularly for complex and heterogeneous sites and emerging research topics. \\
The package is originally implemented in R, with a similar version for Python (\texttt{Reddy4py}).
The package is hosted on GitHub, where contributions and issue reports are explicitly welcomed. 

\section{Conclusion}
We introduced the open-source software package \texttt{Reddy}, which serves as toolbox for EC data processing and analysis, accompanied by a detailed documentation and a set of Jupyter notebooks introducing new users to EC data analysis. Compared to existing software packages, \texttt{Reddy} is deliberately modular: processing, analysis, and visualization steps can be flexibly combined in an iterative ("re‑evaluation") workflow. This makes it straightforward to tailor data processing to site-specific and flow-dependent conditions, and thus to systematically investigate turbulence in heterogeneous environments.
We showcase \texttt{Reddy} by analyzing a case study of stable boundary layer dynamics at an alpine tundra site, for spectral analysis during ice-cover transition at a boreal lake and for fitting flux-variance relations at a palsa mire.

\paragraph{Code and Data Availability}
The \texttt{Reddy} package is available on GitHub (\url{https://github.com/noctiluc3nt/Reddy}) accompanied by a manual and a hands-on introduction to EC data analysis (\url{https://noctiluc3nt.github.io/ec_analyze/}). 
The used terrain data provided by Kartverket is available at \url{https://www.geonorge.no/}.

\paragraph{Acknowledgements}
	The R core team \citep{Rcoreteam2024} is acknowledged.
	This work was supported by the European Research Council (project \#101116083) and is a contribution to the
	strategic research initiative LATICE (Faculty of Mathematics and Natural Sciences, University of Oslo, project \#UiO/GEO103920).

\appendix
\section{Turbulence diagnostics in \texttt{Reddy}}
The here used \texttt{Reddy} diagnostics and functions are summarized in Tab. \ref{tab:diags-app}.
\begin{table} 
	\footnotesize
	\centering
	\caption{Notation, turbulence diagnostics and analysis functions discussed in the text. More diagnostics and functions are described in the \texttt{Reddy} manual.}
	\begin{tabular}{lllll}
		symbol & variable & unit&formula & \texttt{Reddy} function \\
		\hline
		$z$ & measurement height & m \\
		$T$ & temperature & K & \\
		$p$ & pressure & Pa \\
		H$_2$O & water vapor \\
		CO$_2$ & carbon dioxide \\
		CH$_4$ & methane \\
		$(u,v,w)$ & x-, y-, z-wind&m/s&\\
		$U$&wind speed& m/s & &\texttt{calc\_windspeed2D()}\\
		$\theta$ & potential temperature&K&$\theta=T \left( \dfrac{p_0}{p} \right)^{R/c_p}$&\texttt{calc\_theta()}\\[0.1cm]	
		TKE & turbulent kinetic energy & m$^2$/s$^2$ & TKE$=0.5(\sigma_u^2+\sigma_v^2+\sigma_w^2)$& \texttt{calc\_tke()} \\
		$u_*$ & friction velocity&m/s& $u_* = \left( \overline{u'w'}^{2} + \overline{v'w'}^{2} \right)^{1/4}$
		&\texttt{calc\_ustar()}\\
		$L$ & Obukhov length &m&$L = -\,\dfrac{u_*^{3}\,\overline{T}}{\kappa\,g\,\overline{w'T'}}	$ &\texttt{calc\_L()} \\
		$\zeta$ & stability parameter &---& $\zeta = \dfrac{z}{L}$&\texttt{calc\_zeta()}\\
		$N$ & Brunt-Väisälä frequency & 1/s & $N=\sqrt{\dfrac{g}{\theta}\dfrac{\partial \theta}{\partial z}}$&\texttt{calc\_N()} \\[0.1cm]
		$Ri$ &(bulk) Richardson number &---& $Ri=\dfrac{N^2}{(\partial U/\partial z)^2}$&\texttt{calc\_Ri()}\\[0.1cm]
		$\Omega$ &decoupling metric & --- & $\Omega = \dfrac{\sigma_w}{\sqrt{2}zN}$&\texttt{calc\_decoupling\_metric()} \\
		$L_{Oz}$ &Ozmidov scale& m & $L_{Oz} = \sqrt{\dfrac{\epsilon}{N^3}}$& \texttt{calc\_ozmidov\_scale()} \\
		$K_x$ & eddy diffusivity  & m$^2$/s& $K_x = -\dfrac{\overline{w'x'}}{\partial x/\partial z}$& \texttt{calc\_Kh()}, \texttt{calc\_Km()}\\[-0.2cm]
		&(h: heat, m: momentum) \\[0.1cm]
		$Pr$ & Prandtl number & ---& $Pr = K_m/K_h$ & \texttt{calc\_Pr()}\\
		\hline
		$y_B$ &anisotropy&---&(from invariant analysis& \texttt{calc\_anisotropy()}\\
		&&&of Reynolds stress tensor) & \\
		$\epsilon$ & TKE dissipation & m$^2$/s$^3$\\
		SH & sensible heat flux & W/m$^2$ \\
		LH & latent heat flux & W/m$^2$ \\
		R & radiation balance & W/m$^2$ \\
		&(SW: shortwave, LW: longwave) \\
		\hline
		EC & eddy-covariance \\
		MRD & multi-resolution decomposition & & &\texttt{calc\_mrd()}\\
		SEB &surface energy balance& & & \texttt{calc\_seb()} \\
		NEE &net ecosystem exchange&&&\\
		GPP &gross primary production&&&\\
		MOST & Monin-Obukhov similarity theory \\
		LES & large eddy simulation \\
		\hline
		$R$ & gas constant dry air & J/(kg K) & $R=287$ \\
		$c_p$ & specific heat at constant pressure &J/(kg K)& $c_p=1004.5$\\
		$\kappa$ & von-Kármán constant & ---& $\kappa=0.4$ \\
		$g$ & gravitational acceleration & m/s$^2$& $g=9.81$ \\
		\hline
	\end{tabular}
	\label{tab:diags-app}
	
\end{table}

%

\footnotesize
\bibliography{references} 

\end{document}